\documentclass[10pt,conference]{IEEEtran}
\usepackage{graphicx}
\usepackage{cite}

\title{Past, Present and Future: Exploring Adaptive AI in Software Development Bots }

\author{\IEEEauthorblockN{Omar Elsisi, Glaucia Melo} 
\textit{Toronto Metropolitan University}\\
Toronto, Canada \\
\{oelsisy,glaucia\}@torontomu.ca}

\begin{document}
\maketitle

\begin{abstract}
Conversational agents, such as chatbots and virtual assistants, have become essential in software development, boosting productivity, collaboration, and automating various tasks. This paper examines the role of adaptive AI-powered conversational agents in software development, highlighting their ability to offer dynamic, context-aware assistance to developers. Unlike traditional rule-based systems, adaptive AI agents use machine learning and natural language processing to learn from interactions and improve over time, providing more personalized and responsive help. We look at how these tools have evolved from simple query-based systems to advanced AI-driven solutions like GitHub Copilot and Microsoft Teams bots. We also explore the challenges of integrating adaptive AI into software development processes. The study aims to assess the benefits and limitations of these systems, address concerns like data privacy and ethical issues, and offer insights into their future use in the field. Ultimately, adaptive AI chatbots have great potential to revolutionize software development by delivering real-time, customized support and enhancing the efficiency of development cycles.

\end{abstract}


\section{Introduction}

Conversational agents (CAs), including chatbots, dialogue systems, and virtual assistants, are software-based systems designed to process natural language and simulate intelligent dialogue with users \cite{allouch2021conversationaland-NL}. These systems range in complexity, utilizing techniques from simple pre-coded algorithms to sophisticated, adaptive machine learning (ML) algorithms. Generally deployed as service-oriented systems, they are created to help users achieve specific goals based on individual needs by autonomously generating natural language responses that mimic human interactions \cite{lambiase2024motivations}.

Interest in conversational agents has surged in industry and research, yet the concept of human-machine dialogue is not new. Early systems like ELIZA \cite{weizenbaum1983eliza1,weizenbaum1966eliza}, developed at MIT in the mid-1960s, alongside others such as PARRY \cite{colby1974tenparry} and A.L.I.C.E. \cite{wallace2009anatomyALICE}, pioneered the field of Human-Computer Interaction (HCI), examining social and communicative aspects that influence design. Today, advancements in AI, particularly in machine learning, natural language understanding (NLU), natural language generation (NLG), and foundation models, have revitalized interest in conversational agents, highlighting their potential for creating more responsive, context-aware interactions that support a range of applications \cite{kusal2022ai-context-support}. This growing interest underscores the importance of conversational agents in automating and enhancing communication in fields like software engineering, where they can significantly boost productivity and collaboration.

Early conversational agents operated within strict predefined parameters, able only to respond to simple inputs with limited contextual awareness. Today’s CAs leverage NLP to interpret user intent and ML to learn from user interactions, creating a responsive, context-aware experience \cite{milana2024artificial,thorat2020review-rule-base-flexabiltyissue}.

In software development, CAs have become valuable tools for simplifying tasks, minimizing repetitive questions, and boosting developer efficiency \cite{mctear2022conversational-importance-book}. Traditional chatbots, often limited by preset rules and a narrow scope of contextual data, tend to deliver static responses that may fall short in handling complex or fast-changing development needs. Adaptive AI, on the other hand, utilizes machine learning and advanced contextual awareness, enabling these agents to interpret situational factors and past interactions to provide relevant, real-time support. This flexible approach promises to create more responsive workflows, lessening developers' cognitive demands and assisting with intricate tasks such as debugging, code refinement, and project coordination 
\cite{schlimbach2022literatureadabtationorivew}.

In light of these advances, this study explores the impact of adaptive AI-powered chatbots in software development \cite{salammagari2024adaptive11}. We explore how these systems manage complex developer interactions, dynamically adapt to varying contexts, and improve software development workflows. We also investigate the unique strengths of adaptive AI in addressing contextual challenges and its potential applications in education, such as personalized programming instruction and skill enhancement. This research highlights the broader significance of adaptive AI, emphasizing its ability to reshape software development process and learning experiences \cite{schlimbach2022literatureadabtationorivew}.

To address the gap in the current state-of-the-art, this research emphasizes the limited exploration of how adaptive AI-powered conversational agents can dynamically adjust to developers' diverse and complex needs throughout various stages of the software development lifecycle. While much of the existing literature focuses on basic AI chatbot applications or isolated case studies, there is a lack of comprehensive analysis on integrating these systems into real-world development environments, particularly regarding their context-awareness and adaptability to developer workflows. We conducted an ad hoc literature search, reviewing academic databases including Google Scholar, IEEE Xplore, and the ACM Digital Library. We used key search terms such as "adaptive AI in software development" and "conversational agents", along with their synonyms. These sources informed the discussion on how adaptive, context-aware AI is transforming software development from basic query-based systems to advanced, real-time assistance tools.

The contributions of this study lie in its focus on the role of adaptive AI in enhancing developer productivity and collaboration within dynamic development settings. This paper offers a perspective by examining the practical implications of AI tools, including GitHub Copilot, Codex (OpenAI), AlphaCode, and Microsoft Teams bots, which emphasize their ability to learn from interactions and improve over time.

\section{Background}
Conversational agents in software development have evolved significantly, moving from basic, rule-based systems with limited task automation to more sophisticated, AI-driven tools that can assist in complex development tasks. Early chatbots primarily addressed simple queries, such as syntax suggestions or command reminders, which helped developers save time on repetitive tasks. However, the growing need for advanced context-aware support has driven the creation of tools such as GitHub Copilot \cite{copilot24}, Codex (OpenAI)\cite{maddigan2023chat2viscodex}, AlphaCode\cite{li2022competitionalphacode} and Microsoft Teams bots.\cite{mayo2017programmingmicrosoftbots} These tools surpass basic interactions by utilizing AI to deliver dynamic, personalized assistance. They address developers' immediate requirements while seamlessly integrating into workflows, boosting productivity and fostering collaboration within software development environments\cite{copilot24}. GitHub Copilot, for example, leverages machine learning models to offer contextual code suggestions, enabling the prediction of what code a developer might need next based on their recent activity. Microsoft Teams bots, integrated with collaboration platforms, provide task management, document sharing, and even basic code reviews, supporting teams in organizing and progressing on their projects. These tools have set a strong foundation for context-sensitive, AI-driven assistance, demonstrating the growing importance of conversational agents that understand and adapt to a developer’s workflow
\cite{maalej2023rsse-evolution,allouch2021conversational}.

While traditional conversational agents rely on static programming and limited contextual awareness, adaptive AI introduces a new dimension by employing machine learning and real-time data interpretation. Adaptive AI in conversational agents goes beyond pre-programmed responses, dynamically interpreting the developer's environment, task history, and preferences. These agents can identify nuanced patterns and offer recommendations tailored to the developer’s current context, such as recognizing when a developer is likely debugging versus adding new features. Adaptive AI also enables these systems to evolve, learning from past interactions and adapting their suggestions to become progressively more relevant. This shift from static to adaptive conversational agents holds promise for creating a highly personalized, efficient, and responsive development environment 
\cite{melo2023adaptiveemmrrgceadabt}.

\subsection{Types of Chatbots in Software Development}

In Software Development (SD), chatbots are essential for boosting productivity, fostering collaboration, and automating key phases of the Software Development Lifecycle (SDLC). Several types of chatbots are employed in SD, including AI-based, rule-based, generative, and information retrieval. Each type brings distinct advantages and is designed to address specific needs within the SD process. Below, we will examine the four main types of chatbots commonly used in SD and outline their respective functions \cite{thorat2020review-rulebased,dutt2020dynamic-info-retrival-ai,loh2023chatgpt-genrative,hatwar2016ai-based}.

\paragraph{\textbf{AI-Based Chatbots (ML \& NLP-powered)}}
AI-based chatbots leverage machine learning (ML) and natural language processing (NLP) to provide dynamic, context-sensitive support to developers. These chatbots can understand the intent, context, and language of user queries, making them adaptable to a wide range of development tasks and capable of offering intelligent responses based on the developer’s needs \cite{hatwar2016ai-based}. Applications in SD include:
\begin{itemize}
    \item Code Assistance: Provides real-time code suggestions, detects bugs, and helps with debugging.
    \item Task Automation: Automates code deployment, environment management, and version control tasks.
    \item Documentation Generation: Automatically generates relevant documentation based on the project.
    \item Continuous Learning: Learns from interactions to improve responses over time.
\end{itemize}

GitHub Copilot uses machine learning and NLP to assist developers in real-time within their IDE. It suggests code completions, fixes syntax, and generates functions based on natural language comments. It is targeted at developers working on large projects or new technologies and boosts coding efficiency. Copilot is especially helpful for junior developers learning new languages and experienced developers looking to speed up routine tasks, continually improving with open-source code \cite{pudari2023copilot2}.

\paragraph{\textbf{Rule-Based Chatbots}}
Rule-based chatbots follow predefined scripts and set rules to handle user queries. Although less flexible than AI-based chatbots, they are highly effective in automating standard tasks and responding to frequently asked questions (FAQs). They provide reliable and efficient solutions for repetitive queries \cite{thorat2020review-rulebased}. Applications in SD include:
\begin{itemize}
    \item FAQ Automation: Responds to common development-related queries.
    \item Basic Troubleshooting: Provides predefined solutions for common issues.
    \item CI/CD Pipeline Monitoring: Notifies developers about build/test statuses and errors.
\end{itemize}

Slackbot is a rule-based chatbot in Slack that provides quick answers to FAQs or workflows through predefined triggers, ideal for project managers and developers needing consistent responses about processes or configurations. While it does not adapt or learn, it is highly efficient for structured, repetitive tasks like onboarding or retrieving policies \cite{stoeckli2018exploringslackbot}.

\paragraph{\textbf{Generative Chatbots (AI-driven, Advanced Models)}}

Generative chatbots, powered by advanced AI models like GPT, can generate new content dynamically based on user input. These chatbots can create code snippets, documentation, and detailed explanations, making them a powerful tool for software developers at various stages of development \cite{loh2023chatgpt-genrative}. Applications in SD include:
\begin{itemize}
    \item Code Generation: Creates or refactors code based on user input.
    \item Design Assistance: Suggests design patterns or system architectures.
    \item Bug Fixing: Identifies issues in code and suggests fixes.
    \item Real-Time Collaboration: Assists in brainstorming and problem-solving during development.
\end{itemize}

ChatGPT is a generative AI chatbot that creates content dynamically, helping developers with code generation, debugging, documentation, and brainstorming. It’s ideal for software engineers, technical writers, and testers needing creative or exploratory assistance. ChatGPT excels in open-ended tasks like drafting algorithms, coding advice, or user stories, making it invaluable for innovative SD stages 
\cite{beganovic2023methodschatgpt1}.

\paragraph{\textbf{Information Retrieval (IR) Chatbots}}

Information retrieval chatbots focus on searching large datasets, such as codebases, documentation, or issue trackers, to help developers quickly find relevant information. These chatbots are highly effective in retrieving pre-existing data but do not generate new content. They are beneficial for quick access to critical resources \cite{dutt2020dynamic-info-retrival-ai}. Applications in SD include:
\begin{itemize}
    \item Codebase Search: Helps find specific code snippets or functions.
    \item Documentation Search: Locates relevant API references or error codes.
    \item Bug/Issue Tracking: Retrieves past bug reports or ongoing issue statuses.
\end{itemize}

DocChat helps users quickly retrieve information from unstructured documents like manuals or guides. It’s ideal for developers, researchers, and support teams needing instant answers without manual searching. Using NLP and retrieval techniques, it delivers accurate, context-aware responses, streamlining access to critical knowledge \cite{yan2016docchat}.

\section{The evolution of CAs in Software Engineering: Adaptive AI-Powered CAs}

Although traditional chatbots, like rule-based and information retrieval bots, excel at performing specific, predefined tasks, adaptive AI chatbots mark a significant advancement by integrating ML, NLP, and deep learning methods. The main distinction between traditional and adaptive AI chatbots is their capacity to learn, evolve, and manage more intricate and dynamic interactions.

\subsection{Traditional Chatbots (Rule-based \& Information Retrieval Chatbots):}

Rule-based chatbots operate based on a set of predefined scripts or rules. These rules are hardcoded, so the chatbot can only respond to specific questions or commands that the developer has programmed. Information retrieval chatbots, on the other hand, extract responses from a database or knowledge base. While they provide accurate answers as long as the information is available, they don't evolve or improve \cite{binkis2021rule-script}. These chatbots are confined to fixed scenarios or tasks, unable to handle new or unpredictable user inputs that fall outside the predefined rules or knowledge base. For example, a rule-based chatbot in software development might only respond to specific coding queries related to certain programming languages or frameworks \cite{thorat2020review-rule-base-flexabiltyissue}.

Traditional chatbots maintain consistent performance, but any improvements or updates must be made manually. They do not learn from interactions. In the context of SD, for instance, a rule-based chatbot might answer frequently asked questions or retrieve answers from a documentation repository, without evolving based on user interactions 
\cite{thorat2020review-rule-base-flexabiltyissue}.

\subsection{Adaptive AI Chatbots}

Adaptive AI chatbots leverage ML, NLP, and deep learning to learn from user interactions and adapt over time. These bots can process dynamic inputs and adjust their responses based on the context of the conversation. Adaptive AI Chatbots don't depend on static rules or predefined knowledge bases, unlike traditional chatbots. Instead, they generate real-time responses by learning from patterns, user behavior, and feedback \cite{panda2022adapting-ml}. These chatbots offer personalized, context-aware interactions, particularly beneficial in software development, where users may have varying tasks, workflows, and preferences. For example, an adaptive AI chatbot in SD could recognize the developer's coding environment (e.g., Python, Java) and provide tailored solutions. They can handle complex queries, offering debugging support, code suggestions, or troubleshooting advice based on the user's unique project needs 
\cite{shumanov2021makingwith-chatbots-more-personalized}.

For instance, Cursor AI\footnote{cursor.com} serves as a prominent example of adaptive AI in conversational agents for software development, utilizing AI-driven interactions to support developers in coding tasks, debugging, and providing context-aware recommendations. Such tools exemplify the increasing integration of AI in software engineering, with the potential to enhance productivity and improve code quality. 

Adaptive AI chatbots continuously enhance their performance over time by learning from user feedback and adapting to new contexts. In SD, this means the chatbot can improve its code suggestions, recognize patterns in errors, and even anticipate common issues before they occur, making it more effective with each interaction \cite{izadi2024errorcorrection-ai}.

Rather than only responding to queries, adaptive AI chatbots can proactively offer help. They can predict when a developer might need assistance based on past interactions, ongoing project details, or joint issues faced at different stages of development. For example, an adaptive chatbot might suggest optimizations or remind a developer of a testing phase before they even ask for help 
\cite{jiang2023make-more-adabtive}.

\begin{figure}[ht!]
    \centering
    \includegraphics[width=0.8\linewidth]{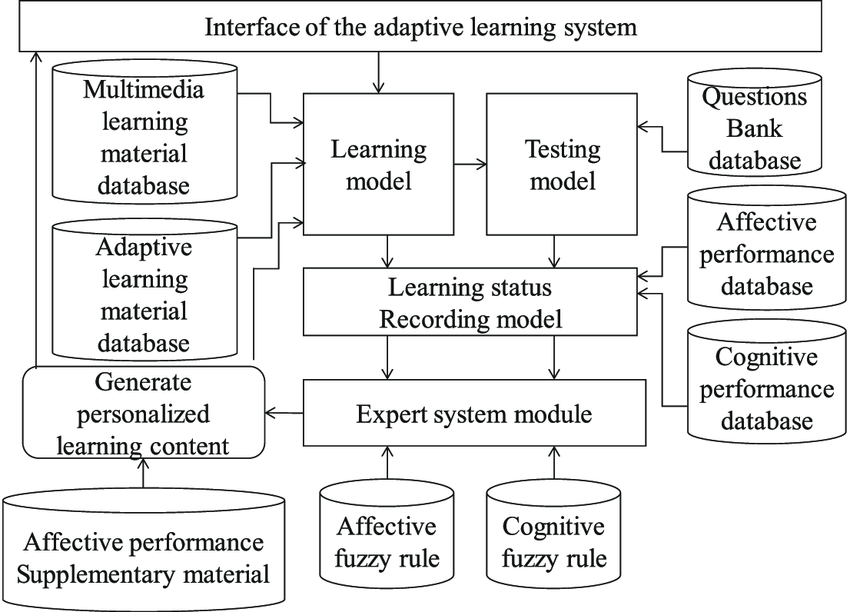}
    \caption{Adaptive learning system with interconnected components \cite{fuzzyexpert2020}.}
    \label{fig:enter-label}
\end{figure}

Figure \ref{fig:enter-label} illustrates the structure of an adaptive learning system with interconnected components. Key boxes include multimedia learning databases, question banks, and cognitive and affective data stores. These elements feed into models for testing, recording learning status, and generating personalized learning content. Arrows indicate the flow of information between these components, highlighting how data on the learner’s performance and preferences is used to tailor content. This dynamic system adapts learning materials in real-time to optimize educational outcomes based on individual needs and progress.

\section{Applications of Adaptive and Generative AI for Context- Learning}

\subsection{Adaptive AI to Generative AI in Programming Education}
Recent recognition of the importance of early programming education highlights its influence on problem-solving skills and future computer science careers \cite{armoni2014early}. Still, high failure rates in introductory courses often stem from difficulties with abstract concepts and programming syntax \cite{medeiros2018systematicIPChighfaiulre}. Traditional intelligent tutoring systems (ITS) have aimed to address this but are limited by their lack of adaptability and research into effectiveness\cite{conati2009intelligentinITS}. Generative AI offers a more dynamic solution, enabling personalized learning experiences by generating tailored content, such as coding exercises and explanations, in real-time based on individual student progress\cite{kwak2023adaptive-from-gerative}. This approach overcomes the rigid limitations of earlier adaptive AI systems by continuously adapting to each student's needs, improving both understanding and performance in programming education \cite{lai2024adapting-from-genrative,milana2024artificial}.

\subsection{How Adaptive AI Deals with Context and Complex Contexts
}
Contextual Awareness in Software Development (SD): In software development, it’s crucial to understand the project scope, developer roles, and task-specific needs to provide accurate, relevant guidance\cite{kung1989conceptual}. Adaptive AI allows chatbots to assess user inputs alongside situational factors, ensuring responses are tailored to the current development context. This capability enables chatbots to offer personalized support based on the context, whether assisting a junior developer with simpler tasks or guiding a senior developer through more complex debugging challenges \cite{melo2023supporting-context-awwarnes}.

Handling Complex Contexts: Adaptive AI chatbots excel in managing complex situations, such as bug fixing or system integration, by utilizing historical data and previous interactions. These chatbots can identify patterns from past tasks, helping them predict the next steps and offer focused recommendations. For example, they might provide relevant code snippets from similar past issues or retrieve documentation for newly introduced APIs, allowing developers to tackle challenges efficiently and in real-time.
\cite{zadeh2023adaptivecontextualaware-adabt,melo2023adaptiveemmrrgceadabt}.

\subsection{Future Insights}

The advancement of adaptive AI-powered conversational agents in software engineering raises several important questions that remain unresolved. These questions are critical for improving system performance, integration, trust, and long-term adoption. Drawing from our analysis of adaptive AI's role in supporting software development and context-sensitive learning, we identify the following key research directions:

\begin{itemize}
    \item \textbf{How is adaptivity defined and implemented in AI-driven developer assistants?}  
    Clarifying the underlying mechanisms of adaptivity such as real-time learning, feedback loops, and contextual awareness will help standardize design practices and evaluation frameworks across tools.

    \item \textbf{What existing adaptive AI tools, agents, or chatbots are currently used in software engineering?}  
    Mapping the current landscape of tools like GitHub Copilot, Codex, Cursor AI, and others will help identify their functional scope and uncover gaps that future tools could fill.

    \item \textbf{What stages of the software development lifecycle (e.g., coding, testing, deployment) are currently supported by adaptive AI tools?}  
    While many tools focus on coding assistance, less attention has been given to areas like integration, deployment, and monitoring. Research should examine where adaptivity is most effective and where it remains underutilized.

    \item \textbf{What is the primary purpose of using adaptive AI in software engineering?}  
    Whether focused on automation, debugging, learning support, code generation, or deployment, clearly understanding the intent behind each tool is essential for evaluating its impact and value.

    \item \textbf{What are the strengths, weaknesses, and limitations of current adaptive AI tools in SE?}  
    Identifying where current tools succeed such as personalized code recommendations and where they struggle such as explainability and integration will inform future improvements.

    \item \textbf{What features or capabilities should future AI-powered tools incorporate to better support developers?}  
    Developers may benefit from enhanced features like proactive suggestions, collaborative memory, integration with development workflows, and better explainability. Future work should explore how to design for these capabilities.

    \item \textbf{How can the privacy and security of sensitive data be ensured when using adaptive AI systems?}  
    Adaptive AI tools often require access to proprietary code and developer interactions. Ensuring secure data handling through encryption, access control, or privacy-preserving learning is essential for widespread adoption.

    \item \textbf{What challenges do adaptive AI systems face in maintaining accuracy and relevance with large-scale or evolving codebases?}  
    Frequent changes in software projects can quickly render learned patterns outdated. Continuous learning pipelines or on-demand model retraining may be needed to maintain recommendation quality.

    \item \textbf{How well do adaptive AI chatbots integrate with tools like code editors, version control, and CI/CD platforms?}  
    Without strong integration into everyday tools, the usefulness of these agents is limited. Research should explore how to embed adaptive AI into the entire software toolchain.

    \item \textbf{What are the risks of over-reliance on adaptive AI for decision-making in development tasks?}  
    Developers may depend too heavily on AI suggestions without fully understanding them, which can lead to quality and security issues. Designing systems that encourage critical thinking and human oversight is necessary.

    \item \textbf{How can adaptive AI support collaboration and knowledge sharing in distributed or remote development teams?}  
    These systems can capture ongoing work, decisions, and patterns, acting as intelligent documentation. Understanding how they can enhance onboarding and team coordination is a promising direction.

    \item \textbf{How dependent is adaptive AI on large datasets, and how does poor data quality affect its performance?}  
    High-quality training data is often hard to obtain. Research is needed on how to build adaptive systems that remain robust even in low-resource or biased-data environments.

    \item \textbf{What ethical concerns arise if adaptive AI reinforces biases or unethical behavior?}  
    AI trained on biased code or user data may unknowingly reproduce or amplify harmful practices. Bias detection and ethical audit mechanisms must be built into development cycles.

    \item \textbf{What are the environmental trade-offs of running large adaptive AI systems?}  
    The computational resources required to train and serve these models are significant. Exploring more energy-efficient models and deployment strategies is important for sustainability.

    \item \textbf{How can AI-generated suggestions be made more explainable and transparent to developers?}  
    Developers often need to understand why a suggestion was made. Including references, reasoning paths, or confidence scores may improve trust and usability.

    \item \textbf{How can developers trust AI results given occasional inaccuracies and opaque decision-making?}  
    Adaptive agents are not always correct. Building trust will require mechanisms like human-in-the-loop workflows, explainability, and clear system limitations.

    \item \textbf{How does the proprietary nature of many AI tools impact vendor lock-in and openness in software engineering?}  
    Many current tools are closed-source, which can restrict transparency, reproducibility, and community innovation. Exploring open alternatives aligns better with open-source development values.
\end{itemize}

\section{Conclusion}
Adaptive AI-powered conversational agents are revolutionizing software development by providing real-time, context-aware assistance that enhances productivity and streamlines workflows. These systems have advanced from basic automation to personalized support tailored to individual developers’ needs and preferences. Leveraging machine learning and natural language processing, they now handle complex tasks like debugging, code generation, and facilitating team collaboration with growing sophistication. This study offers a more comprehensive analysis of their practical use within development environments. Future research should focus on improving these systems’ adaptability, integration with existing tools, and overall performance.

In conclusion, while adaptive AI chatbots are still developing, their ability to offer dynamic, context-aware support makes them a valuable asset for the future of software development. A notable example of this emerging potential is Cursor AI \cite{martinovic2025perceivedcursorai}, which demonstrates how AI-driven conversational agents can enhance coding workflows through intelligent assistance. As tools like Cursor AI continue to evolve, they could serve as a foundation for more advanced adaptive AI solutions, paving the way for greater integration and impact within software engineering.

\bibliographystyle{IEEEtran} 
\bibliography{main.bib}

\end{document}